\documentclass[epj]{svjour}
\usepackage{latexsym}
\usepackage{graphicx}
\usepackage{subfigure}

\newcommand{\mev}{\textrm{ MeV}}

\newcommand{\be}{\begin{equation}}
\newcommand{\ee}{\end{equation}}
\newcommand{\ba}{\begin{eqnarray}}
\newcommand{\ea}{\end{eqnarray}}

\newcommand{\nn}{\nonumber}

\begin{document}

\title{Meson loops in the $f_0(980)$ and $a_0(980)$ radiative decays
into $\rho$, $\omega$}

\author{
 H.~Nagahiro\inst{1}, L.~Roca\inst{2} and E.~Oset\inst{3}}
  
\institute{
Research Center for Nuclear Physics (RCNP), Ibaraki, Osaka
  567-0047, Japan
  \and
Departamento de F\'{\i}sica. Universidad de Murcia. E-30071
Murcia.  Spain \\
and the Yukawa Institute for Theoretical Physics,
Kyoto University, Kyoto 606--8502, Japan
\and
Departamento de F\'{\i}sica Te\'orica and IFIC,
Centro Mixto Universidad de Valencia-CSIC,\\
Institutos de
Investigaci\'on de Paterna, Aptdo. 22085, 46071 Valencia, Spain
}

\abstract{
 We calculate the radiative decay widths of the  $a_0(980)$ and
$f_0(980)$ scalar mesons into $\rho\gamma$ and $\omega\gamma$ 
considering the dynamically generated nature of these
scalar resonances within the realm of the Chiral Unitary Approach.
The main ingredient in the evaluation of the
radiative width of the scalar mesons are the loops coming from
the decay  into their constituent pseudoscalar-pseudoscalar
components and the subsequent radiation of the photon. 
The dominant diagrams with only pseudoscalar mesons in the 
loops are found to be convergent while the divergence of those
 with a vector meson in the loop are written in terms of the two meson
loop function easily regularizable.
We provide results for all the possible charge channels and obtain
results, with uncertainties, which differ significantly from  quark
loops models and some version of vector meson dominance.     
}


\authorrunning{H.~Nagahiro, L.~Roca and E.~Oset}

\titlerunning{Meson loops in the $f_0(980)$ and $a_0(980)$ radiative decays
into $\rho$, $\omega$}

\date{\today}

\maketitle

\section{Introduction} 

The radiative decay of resonances has been advocated as a
privileged tool to study the nature of resonances
\cite{Kalashnikova:2005zz,Wang:2006mf}. A good example of the
interest in these reactions is offered by the large amount of works
devoted to the study of the radiative decay of the 
$D^*_{s0}(2317)$  charmed scalar meson \cite{Dstar}. The uncharmed
scalar mesons are the object of much debate regarding its nature as
a possible $q \bar{q}$ state, a meson-meson molecule, a dynamically
generated object from the interaction of pseudoscalar meson pairs
in coupled channels or a hybrid of those structures 
\cite{vanBeveren:1986ea,Black:1998wt,Tornqvist:1995kr,Oller:1997ti}.  The idea that
the light scalar mesons $\sigma(600)$, $f_0(980)$, $a_0(980)$,
$\kappa(800)$ are bound states or resonances formed from the
interaction of  pseudoscalar mesons has been gaining support,
particularly because the underlying interaction is well known from
the chiral Lagrangians \cite{Gasser:1984gg}, and then using any
reasonable nonperturbative tools, like the Bethe Salpeter equation
\cite{Oller:1997ti,Kaiser:1998fi,Markushin:2000fa,Nieves:1999bx}, the N/D
method  \cite{Oller:1998zr} or the Inverse Amplitude Method
\cite{Oller:1998hw,largenc} the scalar states appear automatically. The
support for this picture is also strengthened by the success
reproducing  different reactions where these resonances are
produced without the need to introduce any free parameters.
Examples of this are the description of the $\phi \to \pi^0 \pi^0
\gamma$ reaction with its strong $f_0(980)$ peak, or the  $\phi \to
\pi^0 \eta \gamma$ reaction with its prominent $a_0(980)$ peak
\cite{Marco:1999df,Markushin:2000fa,Palomar:2003rb}, the $\gamma
\gamma \to \pi \pi$ and  $\gamma \gamma \to K \bar{K}$ reactions
\cite{gamgam,Oller:2007xd} and, 
adjusting some normalization parameters, the
description of the $J/\Psi \to N \bar{N} \pi \pi$ \cite{Li:2003zi}
and the  $J/\Psi \to \omega\pi\pi$, $J/\Psi \to \phi\pi\pi$, with
clear signals for the $\sigma(600)$ in the first reaction and of
$f_0(980)$ in the second one \cite{Meissner:2000bc,Roca:2004uc}.
The apparent narrowness of the $\sigma(600)$ in the   $J/\Psi \to
\omega\pi\pi$ reaction is nicely  interpreted in 
\cite{Roca:2004uc} since the strength of $\sigma$ production is
proportional to the ratio $t/V$, with $t$ the scalar isoscalar $\pi
\pi $ amplitude and $V$ the potential in the same channel (the
Adler term). The narrowness is due to the strong energy dependence
of the Adler term, which grows faster that the scattering
$t$-matrix.  Further considerations concerning the scalar meson 
sector are done in \cite{review}.

  The success of the chiral unitary approach in describing the
scalar sector and in particular the $f_0(980)$ and $a_0(980)$
resonances, can be further tested with new observables like those
we suggest in the present work. Given the fact that the $\phi \to
\pi^0 \pi^0 \gamma$ and   $\phi \to \pi^0 \eta \gamma$ reactions
are well understood within this picture, one can look at the
crossed channel reactions of the former ones in places where the
energy conservation allows it.   This is the case of the $f_0(980)$
and $a_0(980)$ resonances decaying into $ V \gamma$ where $V$ is now a
vector meson which replaces the $\phi$ to make the reaction energetically
possible. Such vector mesons can only be the $\rho$ and the
$\omega$. The sensitivity of this observable to details of
different models is huge and has been discussed in several works,
assuming vector meson dominance and direct coupling of the scalar
to $VV$ \cite{Black:2002ek},
 or quark loop contribution \cite{Kalashnikova:2005zz}
or meson-meson loop contribution, $K \bar{K}$ \cite{Kalashnikova:2005zz}, and
$K \bar{K}$ and $\pi \pi$  \cite{Ivashyn:2007yy}.  The work presented here
will follow the ideas of the chiral unitary approach used to
describe the $\phi \to \pi^0 \pi^0 \gamma$ decay,
evaluating loops of    $K \bar{K}$ and $\pi \pi$, as in the former
works, but also those containing one vector meson, which proved to be
relevant in the study of the radiative decay of vector mesons to
two pseudoscalar mesons and one photon
\cite{hirenpal,Palomar:2003rb}.
An experimental proposal to measure the $f_0$, $a_0$, to $V\gamma$
decays has been approved at COSY/WASA \cite{COSYproposal}.

\section{Formalism}
\label{sec:formalism}

By using the techniques of the unitary extensions of chiral
perturbation theory (chiral unitary approach or Unitarized
ChPT) the low lying scalar resonances ($\sigma$, $\kappa$,
$f_0(980)$ and $a_0(980)$), among many other mesonic and baryonic
resonances, are  generated dynamically. With the only
input of the low-lying chiral Lagrangians, the implementation of
unitarity in coupled channels and the exploitation of the analytic
properties of the scattering amplitudes, the approach generates
poles in unphysical Riemann sheets of the unitarized meson-meson
scattering amplitudes which can be associated to those resonances.
Hence they qualify as quasibound states of their constituent
meson-meson components.  
Furthermore,  by evaluating the residue of the
meson-meson scattering amplitudes at the pole positions,
this picture provides the value, including
the phase, of the couplings of these resonances to their
constituent building blocks. Indeed,
close to the pole position the Laurent expansion of the scattering
amplitude in a particular isospin and partial wave can be
approximated by its dominant term

\begin{equation}
t_{ij}\simeq\frac{g_i g_j}{s-s_p},
\label{eq:couplings}
\end{equation}

\noindent where $i$ and $j$ refer to a given meson-meson estate
and  $s_p\simeq (M-i\Gamma/2)^2$ is the pole position, with $M$ and
$\Gamma$ the mass and width of the associated resonance.  Hence,
$g_i$ in Eq.~(\ref{eq:couplings}) can be identified as the coupling
of the dynamically generated resonance (with the quantum numbers of
the amplitude $t$) to the $i-th$ channel. In particular, using the
Bethe-Salpeter unitarization procedure, we obtain the couplings of
the  $a_0$ and $f_0$ to their constituent pseudoscalar-pseudoscalar
($PP$) components shown in 
table~\ref{tab:couplings}. The difference with the results of
ref.~\cite{Oller:2003vf} will be considered as an uncertainty in
the error analysis in the present work.
\begin{table}[h]
\begin{center}
\begin{tabular}{|c|rc|c|}
\hline 
 $i$ & $g_i$ & $|g_i|$  & $|g_i|$ from ref.~\cite{Oller:2003vf} \\ \hline\hline
$f_0\pi\pi$   & $469-i1247$   & 1332 & 1100   \\ \hline
$f_0 K\bar K$ & $-3369-i1606$ & 3732 & 3680   \\ \hline
$a_0 K \bar K$     & $-4732-i943$  & 4825 & 5500   \\ \hline
$a_0\pi\eta$  & $3166-i712$   & 3245 & 3900   \\ \hline
\end{tabular}
\end{center}
\caption{Couplings of the $f_0(980)$ and $a_0(980)$ to 
their different pseudoscalar-pseudoscalar constituent channels. 
All the units are in MeV.}
\label{tab:couplings}
\end{table}

\begin{figure*}
  \begin{center}
\resizebox{0.7\linewidth}{!}{%
  \includegraphics{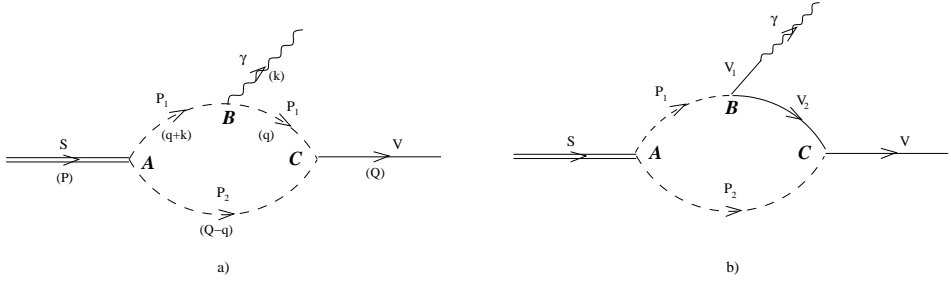}
}
\caption{Feynman diagrams contributing to the scalar radiative
decays}
 \label{fig:diag}
 \end{center}
\end{figure*}

Once we have these couplings, the  philosophy we follow 
in order to obtain the
radiative decay widths is almost straightforward: one has to
consider the transition from the scalar mesons to the possible 
$PP$ at one loop and then attach the photon to the possible allowed
places, considering that a vector meson in the final state
needs to be produced. For reasons that will be clear later on,
the diagrams that we need to evaluate are only those shown in
fig.~\ref{fig:diag}.

The channels we will consider are $f_0\to\rho^0\gamma$,
$f_0\to\omega\gamma$, $a_0^0\to\rho^0\gamma$, $a_0^0\to\omega\gamma$
and $a_0^+\to\rho^+\gamma$. 
In fig.~\ref{fig:diag}, $S$ represents the decaying scalar mesons;
$P_1$, $P_2$, pseudoscalar mesons and $V_1$, $V_2$, vector
mesons. 
In the Appendix, 
tables~\ref{tab:coeff_type_a}-\ref{tab:coeff_type_b_bis},
 we show the different
allowed
$P_1P_2V_1V_2$ particles of the diagrams in fig.~\ref{fig:diag}
together with the corresponding coefficients for each channel to be
explained latter on. In fig.~\ref{fig:diag}, $P$, $q$, $k$ and $Q$
represent the momentum of the different lines that will be used in
 the evaluation of the loop function. 

At this point we take advantage of the
implications of gauge invariance to simplify the calculations.
We follow a
similar procedure as done in 
refs.~\cite{Close:1992ay,Marco:1999df} in the evaluation of the
radiative $\phi$ decay and in refs.~\cite{Roca:2006am,axialshideko}
 for the radiative axial-vector meson decays.

The general expression of the amplitude for 
the radiative decay of a scalar 
meson into  a vector meson and a photon ($S\to V\gamma$)
can be written as 
  \be
 T={\epsilon_V}_\mu \epsilon_\nu T^{\mu\nu}
 \label{eq:Tgeneral}
 \ee 
with
\be
T^{\mu\nu}=a\, g^{\mu\nu} + b\, Q^\mu Q^\nu + c\, Q^\mu k^\nu
 +d\, k^\mu Q^\nu + e\, k^\mu k^\nu
 \label{eq:Tmunuterms}
\ee
where  
 $Q$ is the final vector meson momentum  and $k$
the photon momentum, which are the only independent 
available momenta. 
In Eq.~(\ref{eq:Tgeneral}), $\epsilon_V$ and $\epsilon$ are the
vector  meson and photon polarization vectors respectively. On the
other hand, due to the Lorenz condition, ${\epsilon_V}_\mu
Q^\mu=0$, ${\epsilon}_\nu k^\nu=0$, all the terms in
Eq.~(\ref{eq:Tmunuterms}) vanish except for the $a$ and $d$ terms.
Furthermore, gauge invariance implies that $T^{\mu\nu}k_\nu=0$,
from where one gets 
\be a=-d\,Q\cdot k. 
\label{eq:ad} 
\ee 
Therefore the amplitude gets the general form
 \be
 T=-d(Q.k \,g^{\mu\nu}-k^\mu Q^\nu){\epsilon_V}_\mu
  \epsilon_\nu .
 \label{eq:Tgeneral2}
 \ee 
Hence, we only need to evaluate those diagrams contributing to
the $d$-term, that is, those having a final structure $k^\mu
Q^\nu$. 
  The advantage to evaluate only
the $d$ coefficient is that 
only the loop diagrams of fig.~\ref{fig:diag} contribute
since other diagrams, like those involving photon couplings to the
vertices which are necessary to fulfill gauge invariance, do not give
contribution to the $d$ coefficient
 \cite{Close:1992ay,Marco:1999df,Roca:2006am}.
Another advantage is that,
from dimensional reasons (performing explicitly the Feynman
integrals), we will see that the $d$ coefficients are finite for the
type-a diagrams of fig.~\ref{fig:diag}, while 
the logarithmic divergence 
from the type-b diagram can be easily identified and regularized
 by expressing it in terms of the
 two meson loop function of the $PP$ scattering problem.

With the previous discussion in mind, let us explain the explicit
steps in the calculation of the radiative decay width. We will
consider first the evaluation of the type-a diagram of 
fig.~\ref{fig:diag}.
The Lagrangians needed in the evaluation of the type-a diagram
in fig.~\ref{fig:diag} are, for the vector-pseudoscalar-pseudoscalar
($VPP$)
vertex:

\be
{\cal L}_{VPP}=-i\frac{g}{\sqrt{2}}\langle
V^\mu[\partial_\mu P,P]\rangle
\label{eq:LVPP}
\ee
where $g=-M_V G_V/f^2$, with $M_V$ the vector meson mass and $G_V$ a coupling constant
defined in \cite{Ecker:1988te} and for which we use the numerical
value $G_V=55\pm 5\mev$ from ref.~\cite{Palomar:2003rb}. In
Eq.~(\ref{eq:LVPP}), $V$ and $P$ are the $SU(3)$ matrices containing
the octets of vector and pseudoscalar mesons respectively, see {\it
e.g.} ref.~\cite{Ecker:1988te}, and $\langle\rangle$ means that
one has to perform the trace of the $SU(3)$ resulting matrix.
 In
Eq.~(\ref{eq:LVPP}), $f$ is
the pion decay constant ($f=93\mev$), however one
can assign an uncertainty to the $f$ constant 
since it could range from $f_\pi$ to
$f_\eta$. For the calculations we
will actually use, as central value, $f=1.08 \times 93\mev$
and we will also consider
uncertainties in our calculations
by taking a range of $f$ between $f_\pi$ and $1.15 f_\eta$.

The $PP\gamma$ vertex can be readily obtained from the
 lowest order
ChPT Lagrangian \cite{Gasser:1984gg}
\be
{\cal L}=\frac{f^2}{4}\langle D_\mu U^\dagger D^\mu U\rangle
\ee
from where the $PP\gamma$ amplitude follows: 
\be
-it_{P P\gamma }=-i Q_i\,e\, \epsilon_\mu(p_1+p_2)^\mu
\ee 
with $p_1$($p_2$) the momentum of the ingoing(outgoing)
pseudoscalar meson, $Q_i$ is the sign of 
the charge of
the pseudoscalar meson and
$e$ is taken positive.

Therefore, the amplitude for the type-a diagrams takes the form:

\ba
-it_a&=&-iA\,g_{SP_1P_2} 
\int\frac{d^4q}{(2\pi)^4}\,\frac{i}{(q+k)^2-m_1^2+i\epsilon}
\nn\\
&\times&
\frac{i}{q^2-m_1^2+i\epsilon}\frac{i}{(Q-q)^2-m_2^2+i\epsilon}
\nn\\
&\times& 
(-i)\,e\,Q_1\,\epsilon_\nu(q+q+k)^\nu \nn\\
&\times&(-i)C
\frac{M_V G_V}{\sqrt{2}f^2}
{\epsilon_V}_\mu(q-Q+q)^\mu, 
\label{eq:ita}
\ea

\noindent
where $A$ are coefficients given in table~\ref{tab:coeff_type_a}
needed to relate the $SP_1P_2$ coupling in charge basis with those
in isospin basis and with the unitary normalization of 
refs.~\cite{Oller:1997ti}, $g_{SP_1P_2}$. In Eq.~(\ref{eq:ita}) $Q_1$ is the sign of the 
charge of the
$P_1$ pseudoscalar meson, $m_1$($m_2$) is the mass of the
$P_1$($P_2)$ pseudoscalar meson, $M_V$ the mass of the final vector meson
and $C$ are coefficients coming from the Lagrangian of
Eq.~(\ref{eq:LVPP}) after performing the trace of the matrix and
which depend on the particular $P_1$, $P_2$ and $V$ particles 
(specifically, $C$ is the coefficient of $\langle
V^\mu[\partial_\mu P,P]\rangle$ of Eq.~(\ref{eq:LVPP})
 defined as $C(P_1\partial P_2-P_2\partial P_1)$). The different
 $A$, $C$, coefficients are given in the Appendix, 
table~\ref{tab:coeff_type_a}.

By using the Lorenz condition, ${\epsilon_V}_\mu Q^\mu=0$ and 
$\epsilon_\nu k^\nu=0$, the amplitude reads

\ba
t_a&=&i\,A\,g_{SP_1P_2}\,e\,Q_1\,C 
\frac{M_V G_V}{\sqrt{2}f^2}4\,{\epsilon_V}_\mu \epsilon_\nu \nn \\
&\times&\int\frac{d^4q}{(2\pi)^4}\,q^\mu q^\nu
\frac{1}{(q+k)^2-m_1^2+i\epsilon}\,
\frac{1}{q^2-m_1^2+i\epsilon}\,\nn\\
&\times&\frac{1}{(Q-q)^2-m_2^2+i\epsilon} , 
\label{eq:ita2}
\ea

This loop integral can be easily done by using the Feynman 
parametrization.
We use the identity
\be
\frac{1}{abc}=2\int_0^1 dx\int_0^x dy 
\frac{1}{[a+(b-a)x+(c-b)y]^3}.
\ee 
We set
\ba
a&=&(Q-q)^2-{m_2^2},\nn\\
b&=&q^2-m_1^2,\nn\\
c&=&(q+k)^2-m_1^2
\ea
and perform the change of variable
\be
q=q'+(1-x)Q-yk.
\label{eq:Feyntrans}
\ee
Now we have to recall the above discussion in order to keep in mind
that we only need the $d$ coefficient (see Eqs.~(\ref{eq:Tmunuterms})
and (\ref{eq:Tgeneral2})). That means that we only need the terms
producing a final structure of the type $k^\mu Q^\nu$, which
reduce considerably the number of terms and, more important, the
resulting expression contributing to the $d$-term is just finite.
Thus the apparent superficial logarithmically divergent 
type-a
loop (see Eq.~(\ref{eq:ita})   becomes pure convergent and, hence,
with no need for regularization, it is univoquely
defined  (no regularization parameters, scale, etc).

For the type-a loop,
the final, simple and finite, expression
for the $d$-coefficient of 
Eqs.~(\ref{eq:Tmunuterms})
and (\ref{eq:Tgeneral2}), from where the amplitude can be obtained
by using Eq.~(\ref{eq:Tgeneral2}), is
\be
d_a=AC\,Q_1\,g_{SP_1P_2} \sqrt{2}e\frac{M_V G_V}{f^2}\frac{1}{8\pi^2}
\int_0^1dx\int_0^xdy\frac{(1-x)y}{s+i\varepsilon}
\label{eq:da}
\ee
\noindent
where $s=Q^2x(1-x)+2Q\cdot k(1-x)y+(m_2^2-m_1^2)x-m_2^2$.

In the final step of the derivation of Eq.~(\ref{eq:da}),
since the remaining
$d^4q'$ integration is finite,
 we have used  that
\cite{mandl}
\be
\int d^4q'\frac{1}{(q'^2+s+i\varepsilon)^3}
=i \frac{\pi^2}{2}\frac{1}{s+i\varepsilon}.
\ee
\\

So far, what we have done corresponds to the meson loop calculation
of \cite{Kalashnikova:2005zz,Ivashyn:2007yy} with a significant
difference, which is the coupling of the scalar resonance to the
two pseudoscalar mesons. In \cite{Kalashnikova:2005zz} it is taken
from \cite{Kalashnikova:2004ta} obtained using the Weinberg compositness
condition \cite{Weinberg:1962hj}, and as quoted in 
\cite{Kalashnikova:2005zz,Ivashyn:2007yy} the estimation of the
coupling should be viewed as qualitative [sic]
\cite{Kalashnikova:2005zz}.
Furthermore, only $K\bar K$ loops  are contained in 
\cite{Kalashnikova:2005zz}. In \cite{Ivashyn:2007yy}, $\pi\pi$ loops
are also considered and the couplings of the scalar to the
pseudoscalar mesons are taken from the effective resonance
Lagrangians of \cite{Ecker:1988te}. One should note in this respect
that, while the vectors can be considered as genuine new fields,
additional to the pseudoscalar ones, the scalar fields cannot be
considered at the same level since they come from the same
pseudoscalar-pseudoscalar Lagrangians after a proper unitarization
\cite{Oller:1997ti,Oller:1998zr,Kaiser:1998fi,Markushin:2000fa,Nieves:1999bx}.
This means that the use of a proper unitary theory involving the
pseudoscalar-pseudoscalar meson chiral Lagrangians and the
scalar resonance Lagrangians of \cite{Ecker:1988te}
 leads to double counting.
 On the other hand, assuming that these scalar resonance Lagrangians
 are only a means of providing at tree level the couplings of the
 scalar to pseudoscalar mesons, there are also problems since the
 $a_0$ and $f_0$ properties cannot be simultaneously fitted with the
 structure of the resonance Lagrangians of \cite{Ecker:1988te}.
In our case, as explained above, the couplings are taken from the
residues at the poles of the unitarized meson-meson interaction
amplitudes, which reproduce very accurately the experimental data
in a range of energies from threshold to $1.2$~GeV and are consistent
with a large variety of physical processes.

The other novelty of our approach is the evaluation of the type-b
loops. The idea to include these loops stems from the relevance of
intermediate vector mesons in the radiative decay of $\rho$,
 $\omega$,
into meson-meson-photon \cite{hirenpal}.

For the evaluation of the type-b loops, we first need the
$V\gamma$ vertex given by
\be
t_{V\gamma}=-F_V e \lambda_V M_V \,\epsilon_V\cdot\epsilon_\gamma,
\label{eq:LVg}
\ee
where $\lambda_{V}$ is $1$, $1/3$, $-\sqrt{2}/3$ for $V=\rho$,
 $\omega$,
$\phi$ respectively, $F_V$ is the coupling constant in the
normalization of \cite{Ecker:1988te} for which we use the value
$F_V=156\pm 5\mev$ \cite{Palomar:2003rb}.

For the $VVP$ vertex we use the Lagrangian 
\cite{Bramon:1992kr,Oset:2002sh}:

\be
{\cal L}_{VVP}=\frac{G'}{\sqrt{2}}\epsilon^{\mu\nu\alpha\beta}
\langle\partial_\mu V_\mu\partial_\alpha V_\beta P\rangle,
\label{eq:LVVP}
\ee
where $G'=3g'^2/(4\pi^2f)$ with $g'=-G_V M_\rho /(\sqrt{2}f^2)$.
Since in the type-b loops we have two vertices of the type $VVP$,
the amplitude is proportional to $G'^2$ or $g'^4$. Hence, the
contributions to the decay width of the type-b loops go like $g'^8$.
This means that small differences in the value of $g'$ are
strongly magnified in the evaluation of the radiative decays widths
from the type-b mechanisms. Therefore a good numerical value for
the $VVP$ couplings is called for. 
Hence, in order to fine tune the numerical value of the $VVP$
coupling constant we proceed as follows.
  \begin{figure}[hbt]
\resizebox{0.95\linewidth}{!}{%
  \includegraphics{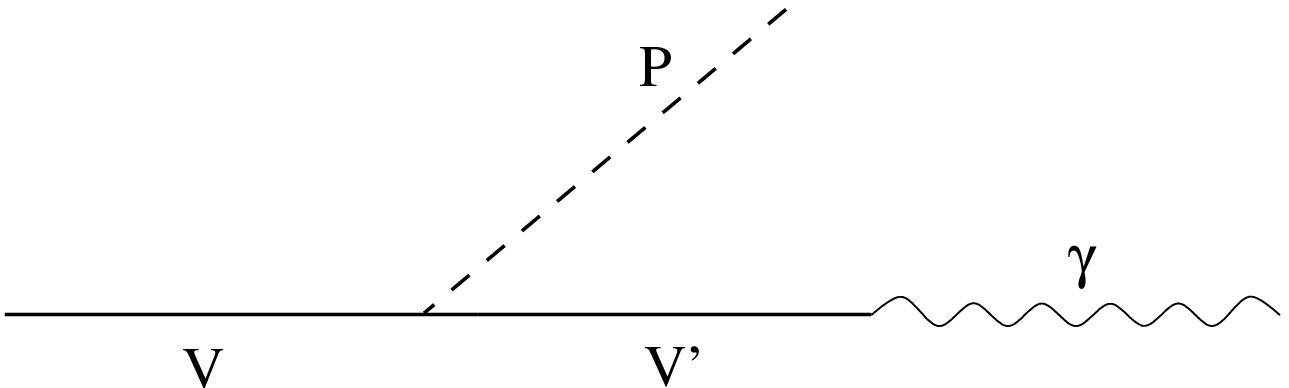}
}
\caption{$V\to P\gamma$ decay diagram}
 \label{fig:VPg}
\end{figure}
From Eqs.~(\ref{eq:LVg}) and (\ref{eq:LVVP}), the decay width of
the decay of a vector meson into a pseudoscalar meson and a
photon, $V\to P\gamma$,
(see fig.~\ref{fig:VPg}),
 takes the form:
\be
\Gamma_{V\to P\gamma}=\frac{e^2}{24\pi}
\left( \frac{B_i G' F_V \lambda_{V'}}{M_{V'}}\right)^2
|\vec k|^3.
\ee
where $B_i$ are numerical coefficients depending on the different
$VVP$ channels that will be explicitly defined later on, and 
$|\vec k|$ is the center of mass momentum of the photon or the
pseudoscalar meson. In table~\ref{tab:BR} we show the theoretical
branching ratios together with the experimental ones
\cite{pdg2007online}.
\begin{table}[htbp]
\begin{center}
\begin{tabular}{|c||c|c|}\hline  
$i$   &$BR_i^{th}$ & $BR_i^{exp}$ \\ \hline \hline  
 $\rho^\pm\pi^\pm\gamma$   & $3.6\times 10^{-4}$ &$(4.5\pm 0.5)\times 10^{-4}$ \\ \hline  
 $\rho^0\pi^0\gamma$    & $3.6\times 10^{-4}$ &$(6.0\pm 0.8)\times 10^{-4}$ \\ \hline  
 $\rho^0\eta\gamma$   &  $3.1\times 10^{-4}$   &  $(2.7\pm 0.4)\times 10^{-4}$ \\ \hline  
 $\omega\pi^0\gamma$  & $6.06$\%  &  $8.91\pm 0.24$\% \\ \hline  
 $\omega\eta\gamma$  & $6.5\times 10^{-4}$  & $(4.8\pm 0.4)\times 10^{-4}$ \\ \hline  
 ${ {K^{*+} \to K^+\gamma} \atop {K^{*-} \to K^-\gamma}}$ 
   & $9.3\times 10^{-4}$  &  $(9.9\pm 0.9)\times 10^{-4}$ \\ \hline  
 ${ {K^{*0}\to K^{0}\gamma} \atop  {\overline K^{*0}\to \overline K^{0}\gamma} }$
  & $19.1\times 10^{-4}$ & $(23.1\pm 2.0)\times 10^{-4}$ \\   
\hline
\end{tabular}
\caption{  Theoretical and experimental branching ratios for different 
vector meson decay processes. }
\end{center}
\label{tab:BR}
 \end{table}
   The agreement with the experimental data is fair but the results
can be improved by incorporating $SU(3)$  breaking mechanisms. For
that purpose, we will normalize here the $G'$ coupling by
multiplying it by $N_i\equiv \sqrt{BR_i^{exp}/BR_i^{th}}$, 
dependent on the particular channel, such that the theoretical
branching ratios  agree with experiment. For the evaluation of the
second column of table~\ref{tab:BR} and in the rest of the paper we
fix $G'=0.098\textrm{ MeV}^{-1}$. The experimental errors in
table~\ref{tab:BR} will be considered later on when evaluating the
uncertainties of our final results for the scalar radiative decays.

The amplitude for the type-b diagram of fig.~\ref{fig:diag} reads

\ba
-it_b&=&-iA\,g_{SP_1P_2}
\int\frac{d^4q}{(2\pi)^4}\,iB\frac{N_BG'}{\sqrt{2}}
F_V M_{V_1} e \lambda_{V_1}\frac{1}{M^2_{V_1}}\nn\\
&\times&
\epsilon^{\mu\nu\alpha\beta}k_\mu\epsilon_\nu q_\alpha
iC\frac{N_CG'}{\sqrt{2}}
\epsilon^{\mu'\nu'\alpha'\beta'}Q_{\mu'} q_{\alpha'}
{\epsilon_{V}}_{\nu'}  \nn\\
&\times& 
\frac{i}{(q+k)^2-m_1^2+i\epsilon}\,
\frac{i}{(Q-q)^2-m_2^2+i\epsilon}\,\nn\\
&\times&
\frac{i}{q^2-M_2^2+i\epsilon}
\left(-g_{\beta\beta'}+\frac{q_\beta q_{\beta'}}{M_2^2}\right)
\label{eq:itb}
\ea
where $M_1$($M_2$) is the mass of the
$V_1$($V_2)$ vector meson.
In Eq.~(\ref{eq:itb}),
$A$ has the same meaning as in the type-a loop case and $B$ is
the coefficient of the $P_1V_1V_2$ vertex obtained after performing
the $SU(3)$ trace in 
$\langle V V P \rangle$ (see Eq.~(\ref{eq:LVVP})) defined as 
$B\,P_1\bar{V_1}\bar{V_2}$. Analogously, $C$ is the coefficient
coming from the $P_2 V_2 V$ vertex
defined as 
$C\,P_2\bar{V} V_2$ from the resulting expression after taking the
trace in $\langle V V P \rangle$.
The $N_B$ and $N_C$ couplings are the normalization factors for the
$B$ and $C$ $VVP$ vertices  just that the $V\to P\gamma$ decays agree with
experiment, as explained above. In the evaluation of the type-b
loops we use for each $VVP$ vertex a value of $N_i$ obtained from
the average of the different charge channels of table~\ref{tab:BR}
containing the same vectors.

The loop integral in Eq.~(\ref{eq:itb}) is apparently
quadratically divergent,  the highest degree of divergence coming
from the last  $q_\beta q_{\beta'}$ term from the 
propagator of the vector meson in the loop.
 However, this
last term
gives zero contribution   since $q_\alpha q_{\alpha'}q_\beta
q_{\beta'}$ is a totally symmetric tensor and vanishes when
contracting it with the antisymmetric Levi-Civita tensor.

Thus, the remaining logarithmically divergent amplitude reads:

\ba
t_b&=&-iA\,g_{SP_1P_2}
B\frac{N_BG'}{\sqrt{2}}
F_V e \lambda_{V_1}\frac{1}{M_{V_1}}
C\frac{N_CG'}{\sqrt{2}}\nn\\
&\times&
\int\frac{d^4q}{(2\pi)^4}\,
\epsilon^{\mu\nu\alpha\beta}
{\epsilon^{\mu'\nu'\alpha'}}_\beta\,
k_\mu q_\alpha
Q_{\mu'} q_{\alpha'}
\,\epsilon_\nu\,{\epsilon_{V}}_{\nu'}  \nn\\
&\times& 
\frac{1}{(q+k)^2-m_1^2+i\epsilon}\,
\frac{1}{(Q-q)^2-m_2^2+i\epsilon}\,\nn\\
&\times&
\frac{1}{q^2-M_2^2+i\epsilon}
\label{eq:itb2}
\ea

Even though the previous expression is logarithmically divergent,
the divergence can be isolated and expressed in terms of the
two-meson loop function that appears in the
unitarization procedure of the 
meson-meson scattering amplitudes.
The reasoning is as follows: by using that
\be
\epsilon^{\mu\nu\alpha\beta}
{\epsilon^{\mu'\nu'\alpha'}}_\beta=
-\left|
\matrix{
g^{\mu \mu'} & g^{\mu \nu'} & g^{\mu \alpha'} \cr
g^{\nu \mu'} & g^{\nu \nu'} & g^{\nu \alpha'} \cr
g^{\alpha \mu'} & g^{\alpha \nu'} & g^{\alpha \alpha'} }
\right|,
\ee
the $\epsilon^{\mu\nu\alpha\beta}
{\epsilon^{\mu'\nu'\alpha'}}_\beta\,
k_\mu q_\alpha
Q_{\mu'} q_{\alpha'}\,\epsilon_\nu\,{\epsilon_{V}}_{\nu'}$ 
term of Eq.~(\ref{eq:itb2}) is reduced to

\ba
&&\epsilon^{\mu\nu\alpha\beta}
{\epsilon^{\mu'\nu'\alpha'}}_\beta\,
k_\mu q_\alpha
Q_{\mu'} q_{\alpha'}
\,\epsilon_\nu\,{\epsilon_{V}}_{\nu'} = \,
-{\epsilon_{V}}_{\mu}\epsilon_\nu\nn \\
&\times&\left\{
  k\cdot Q \,q^2 g^{\mu \nu}
+ q\cdot Q \,k^\mu q^\nu 
+ q\cdot k \,q^\mu Q^\nu \right.\nn\\
&-&\left.
 q\cdot k \,q\cdot Q \,g^{\mu\nu}
- q^2 k^\mu Q^\nu
- k\cdot Q \,q^\mu q^\nu\right\}.
\label{eq:bracket1}
\ea
In order to separate the divergent part in a convenient way, we
write the bracket $\left\{...\right\}$ of  Eq.~(\ref{eq:bracket1}),
by adding and subtracting
 $ q^2 ( k\cdot Q g^{\mu\nu} -
k^\mu Q^\nu )/2$, in
the following way:

\ba
\left\{...\right\}&\equiv& [ 
  k\cdot Q \,q^2 g^{\mu \nu}
+ q\cdot Q \,k^\mu q^\nu 
+ q\cdot k \,q^\mu Q^\nu \nn\\
&-& q\cdot k \,q\cdot Q \,g^{\mu\nu}
- q^2 k^\mu Q^\nu
- k\cdot Q \,q^\mu q^\nu  \nn \\
&-&  \frac{1}{2} k\cdot Q\,q^2 g^{\mu \nu}
+ \frac{1}{2} q^2 k^\mu Q^\nu]\nn\\
&+&\frac{1}{2} q^2 ( k\cdot Q g^{\mu\nu} - k^\mu Q^\nu ).
\label{eq:bracket2}
\ea

Now the square bracket $[...]$ term in Eq.~(\ref{eq:bracket2})
leads to a finite contribution after performing the Feynman
parametrization since the divergent contributions coming from the 
$q'$ terms of Eq.~(\ref{eq:Feyntrans}) cancel algebraically.

The remaining part coming from the last
$ q^2 ( k\cdot Q g^{\mu\nu} -
k^\mu Q^\nu )/2$ term of Eq.~(\ref{eq:bracket2})
leads to a logarithmically divergent contribution which can be
written in terms of the known two-meson loop function,
$G(s,m_1,m_2)$, already used in
the meson-meson unitarized scattering amplitude.
Explicit expressions for $G(s,m_1,m_2)$ can be found, {\it
e.g.}, in refs.~\cite{Oller:2000fj,Roca:2005nm}.
 Indeed, we can do the following
transformation when considering the propagators present
 in Eq.~(\ref{eq:itb2}):

\ba
&&\frac{1}{2} q^2 ( k\cdot Q g^{\mu\nu} -
k^\mu Q^\nu )\frac{1}{q^2-M_2^2}\nn\\
&\times&
\,\frac{1}{(q+k)^2-m_1^2}
\,\frac{1}{(Q-q)^2-m_2^2}=\nn \\
&&
\frac{1}{2} ( k\cdot Q g^{\mu\nu} -
k^\mu Q^\nu )\left(\frac{q^2-M_2^2}{q^2-M_2^2}
+\frac{M_2^2}{q^2-M_2^2}\right)\nn\\
&\times&
\,\frac{1}{(q+k)^2-m_1^2}
\,\frac{1}{(Q-q)^2-m_2^2},
\label{eq:kk1}
\ea
The term $\frac{M_2^2}{q^2-M_2^2}$ in the big parenthesis leads to
a convergent part while the $\frac{q^2-M_2^2}{q^2-M_2^2}$ term is
proportional to the two meson loop function, $G((Q+k)^2)$, which 
can be properly
regularized either with a cutoff \cite{Oller:1997ti} or 
with dimensional regularization \cite{Oller:1998zr,Oller:1998hw}. 
The procedure followed here to write the divergent part in terms of
the meson-meson loop function is similar to what was
 done in ref.~\cite{Napsuciale:2007wp} in a different context.

Therefore, gathering the convergent parts
and keeping only the contributions to the $d-$term
of Eq.~(\ref{eq:Tmunuterms}), we get the following 
convergent
$d$-coefficient:
\ba
d_b^\textrm{con}&=&
-ABC\,g_{SP_1P_2} \frac{N_BN_C G'^2 F_V}{2 M_1}e\lambda_{V_1}
\frac{1}{32\pi^2}\nn\\&\times&
\int_0^1dx\int_0^xdy\frac{1}{s'+i\varepsilon}(Q^2(1-x)^2-M_2^2)
\label{eq:dbconv}
\ea
where now $s'=Q^2x(1-x)+2Q\cdot k(1-x)y+(m_2^2-M_2^2)x+
(M_2^2-m_1^2)y-m_2^2$.

\noindent while the contribution to the $d-$term from the 
divergent part is

\be
d_b^\textrm{div}=
-ABC\,g_{SP_1P_2}\frac{N_BN_C G'^2 F_V}{4 M_1}e\lambda_{V_1}
G(P^2,m_1,m_2).
\label{eq:dbdiv}
\ee

The total amplitude for the radiative decay process is then
obtained from Eq.~(\ref{eq:Tgeneral2}) where 
$d=d_a+d_b^\textrm{con}+d_b^\textrm{div}$ from
Eqs.~(\ref{eq:da}), (\ref{eq:dbconv}) and 
(\ref{eq:dbdiv}).

The radiative decay width of the scalar resonances
 into a vector meson and a photon is then readily obtained,
  taking the
narrow resonance limit in a first step, by
\begin{equation}
\Gamma(M_S,M_V) = \frac{|\vec{k}|}{8\pi M_S^2} \Sigma|T|^2=
\frac{M_s^3}{32\pi}\left(1-\frac{M_V^2}{M_S^2}\right)^3 |d|^2,
\end{equation}
where $M_S$ stands for the mass of the scalar meson.

In order to take into account the finite width of the scalar
resonance and the final vector meson 
we fold the previous expression
with their corresponding mass distributions:
\ba
\Gamma_{S\rightarrow V\gamma} &=& \frac{1}{{\cal N}\pi^2}
\int_{(M_S-2\Gamma_S)^2}^{(M_S+2\Gamma_s)^2}ds_S\,
\int_{(M_V-2\Gamma_V)^2}^{(M_V+2\Gamma_V)^2}
ds_V\,\nn\\
&\times&
Im
\left\{\frac{1}{s_S-M_S^2+iM_S\Gamma_S}\right\}\nn\\
&\times&
Im
\left\{\frac{1}{s_V-M_V^2+iM_V\Gamma_V}\right\} \nn \\
&\times&\Gamma(\sqrt{s_S},\sqrt{s_V})
\,\Theta(\sqrt{s_S}-\sqrt{s_S^{th}})
\Theta(\sqrt{s_V}-\sqrt{s_V^{th}}),\nn\\
\label{eq:convo}
\ea
where $\Theta$ is the step function,
$s_{S(V)}^{th}$ is the threshold for the
dominant $S(V)$ decay channel and ${\cal N}$ is the normalization 
factor of the spectral distributions:
\ba
{\cal N} &=& \frac{1}{\pi^2}
\int_{(M_S-2\Gamma_S)^2}^{(M_S+2\Gamma_s)^2}ds_S\,
\int_{(M_V-2\Gamma_V)^2}^{(M_V+2\Gamma_V)^2}
ds_V\,\nn\\
&\times&
Im
\left\{\frac{1}{s_S-M_S^2+iM_S\Gamma_S}\right\}
Im
\left\{\frac{1}{s_V-M_V^2+iM_V\Gamma_V}\right\} \nn \\
\ea

Advancing some results, the effect of the folding with the
vector-meson spectral function has little influence in the
radiative decay widths. On the contrary, the convolution with the
scalar meson mass distribution is crucial for the decays of 
the $f_0$, as we will explain in the Results section.

\section{Results}

\begin{table*}
\begin{center}
\begin{tabular}{|c| cc|cc||c|} 
\hline
& \multicolumn{2}{r|}{loop a} & \multicolumn{2}{r|}{loop b} & total \\ \hline\hline

$f_0\to\rho^0\gamma$ & \begin{tabular}{r} $\pi\pi$: 0.44 \\ $K\bar K$: 4.10\end{tabular} 
& 3.13  & \begin{tabular}{r} $\pi\pi$: 1.82 \\ $K\bar K$: 0.56\end{tabular} & 1.30 &
$4.2\pm 1.1 $  \\\hline

$f_0\to\omega\gamma$ & $K\bar K$ & 4.31  & \begin{tabular}{r} $\pi\pi$: 1.34 \\ $K\bar K$: 0.013\end{tabular} &
1.49 &
$4.3\pm 1.3 $  \\\hline

$a_0\to\rho\gamma$   & $K\bar K$ & 7.43  & \begin{tabular}{r} $\pi\eta$: 0.68\\ $K\bar{K}$: 0.046 \end{tabular}
& 0.42 &
$11\pm 4 $  \\\hline

$a_0\to\omega\gamma$ & $K\bar K$ & 7.85  & \begin{tabular}{r}  $\pi\eta$: 4.69\\$K\bar K$: 0.79 \end{tabular}
& 9.23 &
$31\pm 13 $  \\\hline

\end{tabular}
\end{center}
\caption{Contribution of the different mechanisms to the radiative
decay widths. All the units are in KeV.}
\label{tab:results1}
\end{table*}

In  table~\ref{tab:results1} we show the contributions of the 
type-a and -b loops to the radiative decay widths under
consideration\footnote{By $a_0\to\rho\gamma$ we denote both
$a_0^0\to\rho^0\gamma$ and $a_0^\pm\to\rho^\pm\gamma$ since the results are
the same for both decays.}.
The numbers besides the labels $\pi\pi$, $K\bar K$ and $\pi\eta$ are 
the decay widths that we would obtain had we used only the loops-a or -b
where the scalar meson vertex is attached to $\pi\pi$, $K\bar K$ and $\pi\eta$
respectively. The other number in columns 2 and 3 is the global loop-a or
loop-b contribution.
The theoretical
errors quoted in our final results  have been obtained by doing a
Monte-Carlo sampling of the parameters of the model within their
uncertainties as have appeared along the text. 
 We have checked
that the largest source of error in the final results for the $f_0$ decays
is the uncertainty considered in the pseudoscalar decay constant $f$, while
the largest one in the $a_0$ decays comes from the uncertainty in 
the $g_{a_0PP}$ couplings.

For the $f_0\to\rho^0\gamma$ the $\pi\pi$ type-a loops
contribution, in spite of being only a 10\% of the $K\bar K$ one, 
influences the global loop-a contribution by about $30$\% due to the
interference with the dominant $K\bar K$ type-a loop.

In  table~\ref{tab:results1} we can see  that the contribution of
the type-b loops to the decay widths, evaluated for the first time
in the present work, is quite important for most of the decays,
particularly for the $f_0\to\rho^0\gamma$, $a_0\to\rho\gamma$ and
$a_0\to\omega\gamma$ decays. The $K\bar K$ loops in the type-b
diagrams are small by themselves with respect  to the $\pi\pi$ or
$\pi\eta$ loops. However, they are important to give the global
type-b contribution after the coherent interference with the
$\pi\pi$ or $\pi\eta$ loops. For instance, had we only considered
the type-a loops, the results would be the same for  the
$a_0\to\rho\gamma$ and $a_0\to\omega\gamma$ radiative decay widths.
(The differences in table~\ref{tab:results1} 
are due to the different  masses of $\rho$ and
$\omega$).  The type-b loops for these $a_0$ decays are
dominated by the $\pi\eta$ loops. However, the $\pi\eta$ mechanisms
of the type-b loops for the $a_0\to\omega\gamma$ decay is one order
of magnitude larger than that of the  $a_0\to\rho\gamma$. This is
essentially due to the fact that the vector meson attached to the
photon ($V_1$ in fig.~\ref{fig:diag}b) is a $\rho$ in the former
case  and an $\omega$ in the later one. This implies a factor 
three  difference in the $\lambda_{V_1}$ coefficient of the direct 
coupling of the vector meson to the photon (see
Eq.~(\ref{eq:LVg})). This makes, after the interference with the
other pieces, that the width of  the $a_0\to\omega\gamma$ decay is
much larger than the   $a_0\to\rho\gamma$ decay width. 

Since all these interference effects are so important, it is crucial to know 
the relative sign 
among the different couplings of the model.
At this point the chiral unitary approach makes a decisive contribution
since, as explained at the beginning of section~\ref{sec:formalism},
it provides the $g_{SPP}$ couplings including their phase. Hence the
right direction of the interferences among the different mechanisms
is a genuine non-trivial prediction of UChPT.

On the other hand it is worth mentioning the importance of the
convolution with the scalar mass distribution, Eq.~(\ref{eq:convo}).
Had we not considered the folding of the decay width with the 
scalar meson spectral function
 we would have obtained radiative 
decay widths for the $f_0$ about $3$ times smaller than in the case
when the convolution is performed.
Indeed, as already mentioned in
 refs.~\cite{Kalashnikova:2005zz,Ivashyn:2007yy}
the loop functions for the type-a diagrams are
strongly dependent on the scalar meson mass.
This is specially relevant for the $f_0$ decay channels.
\begin{figure*}
     \centering
     \subfigure[]{
          \label{fig:G_vs_MS_a}
          \includegraphics[width=.4\linewidth]{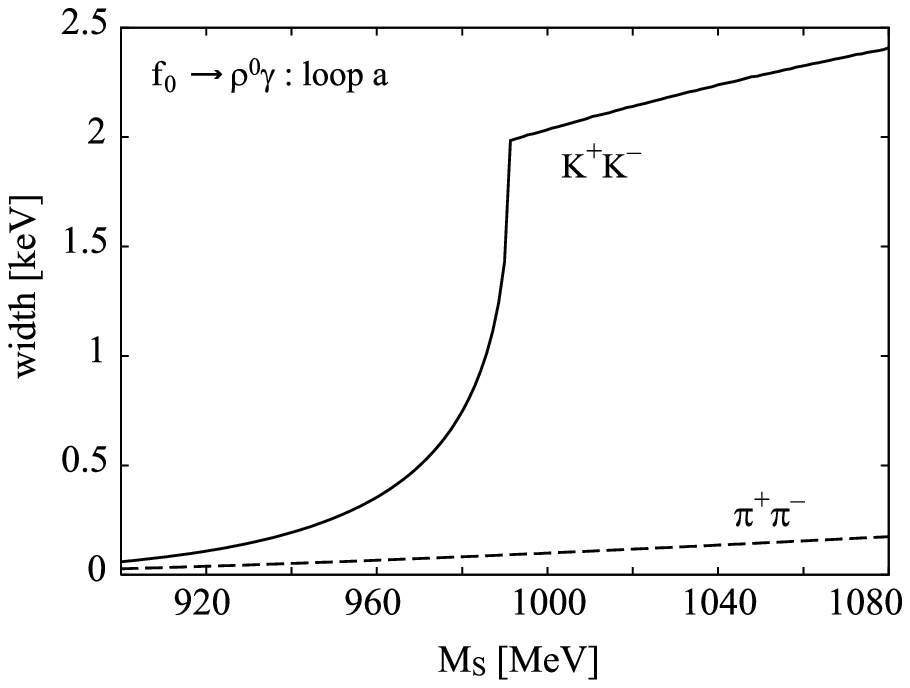}}
     \subfigure[]{
          \label{fig:G_vs_MS_b}
          \includegraphics[width=.4\linewidth]{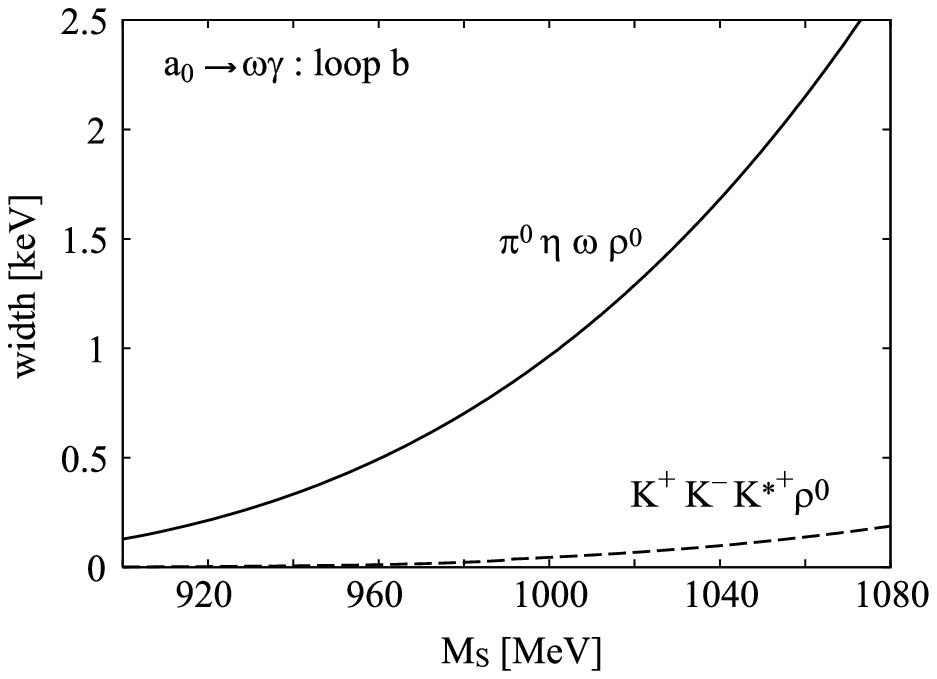}}
     \caption{Left panel: $f_0\to \rho^0\gamma$ decay width as a function of
      the scalar meson mass
     for the type-a loop. Solid line: $\pi^+\pi^-$ loop; 
     dashed line: $K^+K^-$
     loop. Right panel: $a_0\to \omega\gamma$ decay width as a function
      of the scalar meson mass
     for the type-b loop.      Solid line: $\pi^0\eta\omega\rho^0$ loop;
      dashed line: 
     $K^+K^-{K^*}^+\rho^0$ loop.   }
     \label{fig:G_vs_MS}
\end{figure*}

In fig.~\ref{fig:G_vs_MS} we show, as an example,  the decay widths
as a function of the scalar meson mass that one would obtain
(without the convolution) for particular mechanisms for the type-a 
(fig.~\ref{fig:G_vs_MS_a}) and type-b (fig.~\ref{fig:G_vs_MS_b})
mechanisms. In fig.~\ref{fig:G_vs_MS_a} one can see the strong
dependence with the energy of the type-a kaon loops close to the
$K\bar K$ threshold. This makes that small variations in the scalar
meson mass change drastically the final radiative decay
width. This is why it is very important to account  properly for
the scalar meson mass distribution by  folding with the spectral
function. For the type-a $\pi\pi$ loops and for the type-b, the
dependence with the energy are smoother  because, as seen in
table~\ref{tab:results1}, they are dominated by $\pi\pi$ or
$\pi\eta$ $P_1P_2$ mechanisms, which have the threshold 
 far from the
$\sim 980\mev$ region of interest.  Therefore, the influence of the
convolution with the scalar mass distribution is less relevant for
the type-b loops.

At this point it is worth commenting about the relative size of the 
divergent and convergent part of the type-b loops. We have checked
that by using  in the evaluation of the decay widths only the
convergent (Eq.~\ref{eq:dbconv}) or divergent (Eq.~\ref{eq:dbdiv})
part, the contribution from the type-b loops would be from 3 to 10
times larger  (depending on the channels) than those given in
table~\ref{tab:results1}. However, the sign of $d_b^\textrm{div}$
is opposite to that of  $d_b^\textrm{conv}$ and then a strong
destructive interference results leading to the type-b contributions
shown in table~\ref{tab:results1}.

Finally, in table~\ref{tab:results_others} we compare our result with other
theoretical determinations.
The model \cite{Kalashnikova:2005zz}(I) is based on quark models,
ref.~\cite{Kalashnikova:2005zz}(II) on kaon loops,
ref.~\cite{Black:2002ek}(I) and (II) on VMD from direct SVP contact
term and ref.~\cite{Ivashyn:2007yy} on meson loops by using the 
resonance Lagrangians of \cite{Ecker:1988te}.

\begin{table*}
\begin{center}
\begin{tabular}{|c|c|c|c|c|c||c|} 
\hline
 &\cite{Kalashnikova:2005zz}(I) &\cite{Kalashnikova:2005zz}(II) &\cite{Black:2002ek}(I) 
 &\cite{Black:2002ek}(II) & \cite{Ivashyn:2007yy}& present work\\ \hline\hline
$f_0\to\rho^0\gamma$ & 125 & 3 & $19\pm 5$    & $3.3\pm2.0$  &  9.6 & $4.2\pm 1.1 $ \\\hline
$f_0\to\omega\gamma$ & 14  & 3 & $126\pm 20$  & $88\pm 17$   & 15.0 & $4.3\pm 1.3$ \\\hline
$a_0\to\rho\gamma$   & 14  & 3 & $3.0\pm 1.0$ & $3.0\pm 1.0$ &  9.1 & $11\pm 4$    \\\hline
$a_0\to\omega\gamma$ & 125 & 3 & $641\pm 87$  & $641\pm 87$  &  8.7 & $31\pm 13$   \\\hline
\end{tabular}
\end{center}
\caption{Comparison of the radiative decay widths 
with other theoretical predictions. All the units are in KeV.}
\label{tab:results_others}
\end{table*}

We can see the wide dispersion among the different theoretical
models.  The previous calculations dealing with meson loops,
refs.~\cite{Kalashnikova:2005zz}(II) and \cite{Ivashyn:2007yy}, do
not include the type-b loops considered in the present work. The
model  of ref.~\cite{Kalashnikova:2005zz}(II) neglects the pion
loops for the type-a mechanisms which, as shown above and despite
being small by themselves, affects the final result of the
radiative decay widths  in a significant manner after the coherent
addition to the dominant contributions. On the other hand, the
model \cite{Kalashnikova:2005zz}(II) relies on an estimation of
the $g_{SPP}$ that {\it should be viewed as qualitative} [sic]
\cite{Kalashnikova:2005zz}
and the model of  \cite{Ivashyn:2007yy} uses couplings from the
resonance Lagrangian of \cite{Ecker:1988te} which we questioned
before.
 On the contrary, the chiral unitary approach provides precise
 values for the couplings, including their phase, which have been
 successfully tested in many processes, and have uncertainties under
 control. 
 In any case we should stress that all the results from the
meson loops are of the same order of magnitude, and quite different
from the quark and VMD models.

The large dispersion among the theoretical models \linebreak
stresses
the extreme sensitivity of these radiative
decays to the theoretical models and on the nature of these
resonances. Hence, an experimental measurement would be very
valuable to discern among theoretical models. 
The numerical values obtained here are within reach in present
experimental facilities, in particular at COSY/Juelich where an
experiment is already approved \cite{COSYproposal}.


%

\section{Summary}

We have evaluated the radiative decays of the $a_0(980)$ and
$f_0(980)$ scalar mesons into a vector meson and one photon
from the point of view of the chiral unitary approach, where
the $a_0(980)$ and $f_0(980)$ scalar mesons are dynamically
generated by implementing unitarity in coupled channels in the
pseudoscalar-pseudoscalar interaction.
 By evaluating the residues at the pole
positions in unphysical Riemann sheets of the partial wave
amplitudes, the couplings of the scalar mesons to the different $PP$
channels can be obtained both in modulus and  phase. Within this
dynamical framework, the natural way to evaluate the radiative decay
widths is to consider the decay of the scalar mesons into their
constituent $PP$ building blocks and then allow the photon to be
emitted from the pseudoscalar legs of the allowed loops.
We consider not just the mechanisms with only pseudoscalar mesons
in the loops, but also those with a vector meson in the loop.
By using arguments of gauge invariance we show that the
contribution of the loops involving only pseudoscalar mesons, which
are in most of the cases the dominant ones, are
convergent. The loops containing a vector meson  are
logarithmically divergent, but this divergence can be recast into
the  two meson-loop function
 used in the scattering process, well under control,
which leads to the dynamical generation of the scalar resonances.

We make predictions for all the allowed 
$f_0/a_0\to V\gamma$  decay widths including also an error analysis
from the uncertainties in the parameters of the model.
We show
that  even if some of the mechanisms are small by
themselves, like the pion loops for the $f_0\to\rho^0\gamma$ or the
loops containing a vector meson in the loop, they affect 
strongly the final results due to non-trivial
interferences with the dominant mechanisms. The sign of these
interferences are well under control
thanks to the knowledge of the phase of the 
couplings provided by the  underlying
unitary theory that generates dynamically the scalar
resonances. 

These radiative decays are very sensitive  to
details of the models, hence an 
experimental measure would be highly valuable to discern among
theoretical models.
The COSY/WASA \cite{COSYproposal} 
scheduled experiment should be very useful in this
respect.

\section*{Acknowledgments}

We thank financial support from MEC (Spain) grants No. FPA2004-03470,
FIS2006-03438, FPA2007-62777, Fundaci\'on S\'eneca grant No. 02975/PI/05,
and the Japan(JSPS)-Spain collaboration agreement.
One of the author (H.N.) is
the Research Fellow of the Japan Society for the Promotion of Science
(JSPS) and supported by the Grant for Scientific Research from JSPS
(No.~18$\cdot$8661). 
This research is  part of the EU Integrated
Infrastructure Initiative Hadron Physics Project under contract
number RII3-CT-2004-506078.

\appendix

\section{Numerical coefficients}
\label{appendix}

\begin{table*}[h]
\begin{center}
\begin{tabular}{c|c|c|c} 
decay & $P_1P_2$ & $A$ & $C$ \\\hline\hline
$f_0\to\rho_0\gamma$ & $\pi^+\pi^-$ & $-\sqrt{2/3}$ & $-\sqrt{2}$  \\
                     & $\pi^-\pi^+$ & $-\sqrt{2/3}$ & $ \sqrt{2}$  \\
                     & $K^+K^-$     & $-1/\sqrt{2}$ & $-1/\sqrt{2}$  \\
		     & $K^-K^+$     & $-1/\sqrt{2}$ & $ 1/\sqrt{2}$  \\ \hline
$f_0\to\omega\gamma$ & $K^+K^-$     & $-1/\sqrt{2}$ & $-1/\sqrt{2}$  \\
		     & $K^-K^+$     & $-1/\sqrt{2}$ & $ 1/\sqrt{2}$  \\ \hline		     
$a_0^0\to\rho_0\gamma$ & $K^+K^-$   & $-1/\sqrt{2}$ & $-1/\sqrt{2}$  \\
		     & $K^-K^+$     & $-1/\sqrt{2}$ & $ 1/\sqrt{2}$  \\ \hline		     
$a_0^0\to\omega\gamma$ & $K^+K^-$   & $-1/\sqrt{2}$ & $-1/\sqrt{2}$  \\
		     & $K^-K^+$     & $-1/\sqrt{2}$ & $ 1/\sqrt{2}$  \\ \hline		     
$a_0^+\to\rho^+\gamma$ & $K^+\overline{K^0}$& $-1$  & $-1$  \\
\end{tabular}
\end{center}
\caption{Coefficients of Eq.~(\ref{eq:da}) for type-a diagrams}
\label{tab:coeff_type_a}
\end{table*}

\begin{table*}[htb]
\begin{center}
\begin{tabular}{c|r|c|c|c} 
decay & $P_1P_2V_2V_1$ & $A$ & $B$ & $C$ \\\hline\hline
$f_0\to\rho_0\gamma$ & $\pi^0\pi^0\omega\rho^0$ & $-\sqrt{2/3}$ & $\sqrt{2}$ & $\sqrt{2}$ \\ \cline{2-5}
                     & $K^+K^-K^{*+}\rho^0$     & $-1/\sqrt{2}$ & $1/\sqrt{2}$ & $1/\sqrt{2}$ \\
                     & $\omega$                 & $$            & $1/\sqrt{2}$ & $$ \\		     
                     & $\phi$                   & $$            & $1$          & $$ \\	\cline{2-5}		     
                     & $K^-K^+K^{*-}\rho^0$     & $-1/\sqrt{2}$ & $1/\sqrt{2}$ & $1/\sqrt{2}$ \\
                     & $\omega$                 & $$            & $1/\sqrt{2}$ & $$ \\		     
                     & $\phi$                   & $$            & $1$          & $$ \\	\cline{2-5}		     
                     & $K^0\bar{K^0}K^{*0}\rho^0$& $-1/\sqrt{2}$ & $-1/\sqrt{2}$ & $-1/\sqrt{2}$ \\
                     & $\omega$                 & $$            & $1/\sqrt{2}$ & $$ \\		     
                     & $\phi$                   & $$            & $1$          & $$ \\	\cline{2-5}		     
                     & $\bar{K^0}K^0\bar{K}^{*0}\rho^0$& $-1/\sqrt{2}$ & $-1/\sqrt{2}$ & $-1/\sqrt{2}$ \\
                     & $\omega$                 & $$            & $1/\sqrt{2}$ & $$ \\		     
                     & $\phi$                   & $$            & $1$          & $$ \\	\hline		     
$f_0\to\omega\gamma$ & $\pi^+\pi^-\rho^+\omega$ & $-\sqrt{2/3}$ & $\sqrt{2}$ & $\sqrt{2}$ \\ \cline{2-5}
                     & $\pi^-\pi^+\rho^-\omega$ & $-\sqrt{2/3}$ & $\sqrt{2}$ & $\sqrt{2}$ \\ \cline{2-5}	
                     & $\pi^0\pi^0\rho^0\omega$ & $-\sqrt{2/3}$ & $\sqrt{2}$ & $\sqrt{2}$ \\ \cline{2-5}
                     & $K^+K^-K^{*+}\rho^0$     & $-1/\sqrt{2}$ & $1/\sqrt{2}$ & $1/\sqrt{2}$ \\		     
                     & $\omega$                 & $$            & $1/\sqrt{2}$ & $$ \\		     
                     & $\phi$                   & $$            & $1$          & $$ \\	\cline{2-5}		     
                     & $K^-K^+K^{*-}\rho^0$     & $-1/\sqrt{2}$ & $1/\sqrt{2}$ & $1/\sqrt{2}$ \\
                     & $\omega$                 & $$            & $1/\sqrt{2}$ & $$ \\		     
                     & $\phi$                   & $$            & $1$          & $$ \\	\cline{2-5}		     
                     & $K^0\bar{K^0}K^{*0}\rho^0$& $-1/\sqrt{2}$ & $-1/\sqrt{2}$ & $1/\sqrt{2}$ \\
                     & $\omega$                 & $$            & $1/\sqrt{2}$ & $$ \\		     
                     & $\phi$                   & $$            & $1$          & $$ \\	\cline{2-5}		     
                     & $\bar{K^0}K^0\bar{K}^{*0}\rho^0$& $-1/\sqrt{2}$ & $-1/\sqrt{2}$ & $1/\sqrt{2}$ \\
                     & $\omega$                 & $$            & $1/\sqrt{2}$ & $$ \\		     
                     & $\phi$                   & $$            & $1$          & $$ \\	\hline

\end{tabular}
\end{center}
\caption{Coefficients $A$, $B$ and $C$ 
of Eq.~(\ref{eq:itb}) for the different allowed type-b diagrams.
The continuation is in table~\ref{tab:coeff_type_b_bis}...}
\label{tab:coeff_type_b}
\end{table*}

\begin{table*}[htb]
\begin{center}
\begin{tabular}{c|r|c|c|c} 
decay & $P_1P_2V_2V_1$ & $A$ & $B$ & $C$ \\\hline\hline
$a_0^0\to\rho_0\gamma$ &$K^+K^-K^{*+}\rho^0$     & $-1/\sqrt{2}$ & $1/\sqrt{2}$ & $1/\sqrt{2}$ \\		     
                     & $\omega$                 & $$            & $1/\sqrt{2}$ & $$ \\		     
                     & $\phi$                   & $$            & $1$          & $$ \\	\cline{2-5}		     
                     & $K^-K^+K^{*-}\rho^0$     & $-1/\sqrt{2}$ & $1/\sqrt{2}$ & $1/\sqrt{2}$ \\
                     & $\omega$                 & $$            & $1/\sqrt{2}$ & $$ \\		     
                     & $\phi$                   & $$            & $1$          & $$ \\	\cline{2-5}		     
                     & $K^0\bar{K^0}K^{*0}\rho^0$& $1/\sqrt{2}$ & $-1/\sqrt{2}$ & $-1/\sqrt{2}$ \\
                     & $\omega$                 & $$            & $1/\sqrt{2}$ & $$ \\		     
                     & $\phi$                   & $$            & $1$          & $$ \\	\cline{2-5}		     
                     & $\bar{K^0}K^0\bar{K}^{*0}\rho^0$& $1/\sqrt{2}$ & $-1/\sqrt{2}$ & $-1/\sqrt{2}$ \\
                     & $\omega$                 & $$            & $1/\sqrt{2}$ & $$ \\		     
                     & $\phi$                   & $$            & $1$          & $$ \\	\cline{2-5}	
	             & $\pi^0\eta\rho^0\omega$        & $1$           & $\sqrt{2}$   & $\sqrt{2/3}$ \\	\cline{2-5}	
		     & $\eta\pi^0\omega\omega$        & $1$           & $\sqrt{2/3}$   & $\sqrt{2}$ \\	\cline{1-5}
$a_0^0\to\omega\gamma$ &$K^+K^-K^{*+}\rho^0$     & $-1/\sqrt{2}$ & $1/\sqrt{2}$ & $1/\sqrt{2}$ \\		     
                     & $\omega$                 & $$            & $1/\sqrt{2}$ & $$ \\		     
                     & $\phi$                   & $$            & $1$          & $$ \\	\cline{2-5}		     
                     & $K^-K^+K^{*-}\rho^0$     & $-1/\sqrt{2}$ & $1/\sqrt{2}$ & $1/\sqrt{2}$ \\
                     & $\omega$                 & $$            & $1/\sqrt{2}$ & $$ \\		     
                     & $\phi$                   & $$            & $1$          & $$ \\	\cline{2-5}		     
                     & $K^0\bar{K^0}K^{*0}\rho^0$& $1/\sqrt{2}$ & $-1/\sqrt{2}$ & $1/\sqrt{2}$ \\
                     & $\omega$                 & $$            & $1/\sqrt{2}$ & $$ \\		     
                     & $\phi$                   & $$            & $1$          & $$ \\	\cline{2-5}		     
                     & $\bar{K^0}K^0\bar{K}^{*0}\rho^0$& $1/\sqrt{2}$ & $-1/\sqrt{2}$ & $1/\sqrt{2}$ \\
                     & $\omega$                 & $$            & $1/\sqrt{2}$ & $$ \\		     
                     & $\phi$                   & $$            & $1$          & $$ \\	\cline{2-5}	
	             & $\pi^0\eta\omega\rho^0$        & $1$           & $\sqrt{2}$   & $\sqrt{2/3}$ \\	\cline{2-5}	
		     & $\eta\pi^0\rho^0\rho^0$        & $1$           & $\sqrt{2/3}$   & $\sqrt{2}$ \\ \hline		     
$a_0^+\to\rho^+\gamma$ & $K^+\bar{K^0}K^{*+}\rho^0$& $-1$ & $1/\sqrt{2}$ & $1$ \\
                     & $\omega$                 & $$            & $1/\sqrt{2}$ & $$ \\		     
                     & $\phi$                   & $$            & $1$          & $$ \\	\cline{2-5}		     
                     & $\bar{K^0}K^+\bar{K}^{*0}\rho^0$& $-1$ & $-1/\sqrt{2}$ & $1$ \\
                     & $\omega$                 & $$            & $1/\sqrt{2}$ & $$ \\		     
                     & $\phi$                   & $$            & $1$          & $$ \\	\cline{2-5}	
	             & $\pi^+\eta\rho^+\omega$        & $1$           & $\sqrt{2}$   & $\sqrt{2/3}$ \\	\cline{2-5}	
		     & $\eta\pi^+\omega\omega$        & $1$           & $\sqrt{2/3}$   & $\sqrt{2}$ \\ \hline
\end{tabular}
\end{center}
\caption{... continuation of table~\ref{tab:coeff_type_b}.}
\label{tab:coeff_type_b_bis}
\end{table*}

\end{document}